\documentclass[twocolumn,showpacs,prl,aps]{revtex4}
\usepackage{graphicx,color}
\usepackage{dcolumn}
\usepackage{bm}

\newcommand{\bra}[1]{\left\langle {#1} \right|}
\newcommand{\ket}[1]{\left| {#1} \right\rangle}

\newcommand{\expect}[1]
{\left\langle {#1}  \right\rangle}

\begin{document}

\title{Dynamics of a domain wall and spin-wave excitations driven by a
  mesoscopic current}

\author{Jun-ichiro Ohe and Bernhard Kramer}

\affiliation{%
I. Institut f\"{u}r Theoretische Physik, Universtit\"{a}t Hamburg,
Jungiusstrasse 9, 20355 Hamburg, Germany}
\date{\today}

\begin{abstract}
  The dynamics of a domain wall driven by a spin-polarized current in a
  mesoscopic system is studied numerically. Spin-mixing in the states of the
  conduction electrons is fully taken into account. When the Fermi energy of
  the electrons is larger than the exchange energy ($E_{\rm F}>J_{\rm
    sd}$), the spin precession induces spin wave excitations in the local
  spins which contribute towards the displacement of the domain wall. The
  resulting average velocity is found to be much smaller than the one obtained
  in the adiabatic limit. For $E_{\rm F}<J_{\rm sd}$, the results are
  consistent with the adiabatic approximation except for the region below the
  critical current where a residual domain wall velocity is found.
\end{abstract}

\pacs{75.75.+a, 72.25.Pn}

\maketitle

The interplay between the dynamics of conduction electrons and the local spins
in magnetic multi-layers has been pioneered by Berger \cite{Berger86,Berger96}
and Slonczewski \cite{Slonczewski}. They pointed out that spin flips of
conduction electrons induce a spin torque acting on the interfaces between
multilayers. This spin torque can excite spin waves propagating in the
interface perpendicular to the current direction. In the case of a smooth
domain wall, as a result of a current flow, the spin torque can lead to domain
wall motion opposite to the current direction. The spin torque can also induce
an out-of-plane component of the local spins if damping is present. In
addition, the reflection of the conduction electrons can lead to a reaction
force acting on a domain wall via a momentum transfer. This can also lead to a
displacement \cite{Berger86}. 

During the past decade, considerable progresses have been made in
understanding the current-induced domain wall motion in magnetic nanowires
\cite{Li04a,thiaville,Li04b,Tatara,waintal,Barnes}. 
This was motivated by remarkable
experimental successes measuring the motion of a domain wall under the
influence of a current pulse. Velocities of about 3\,m/s have been found which
exhibit large fluctuations around the average value.
\cite{Yamaguchi,Yamanouchi,Saitoh}. The velocity of a domain wall induced only
by the spin torque has been studied \cite{Li04a,Li04b}. This is suitable in
the adiabatic limit, which is assumed to be a good approximation for
ferromagnetic nanowires. In this limit, the thickness of the domain wall is
assumed to be much larger than any of the characteristic length scales of the
conduction electrons. The momentum transfer is then very small due to the
absence of backscattering. A velocity of 250\,m/s has been estimated in lowest
order of the interaction between the electrons and the wall.

In the adiabatic limit, a finite final velocity of the domain wall can only be
sustained by the presence of an additional external magnetic field
\cite{Li04a} and/or by lowest-order non-adiabatic corrections towards the
shape of the magnetization profile \cite{thiaville,Li04b}. The effects of
these corrections reduce the domain wall velocity at least by one order of
magnitude. The current-driven dynamics of a domain wall including
contributions from non-adiabaticity, has been investigated microscopically by
treating the conduction electrons quantum mechanically and including both,
spin and momentum transfers \cite{Tatara}. Spin wave generation has been
neglected. A parameter-dependent critical current has been found below which
the domain wall is pinned, and which vanishes for an abrupt wall.

In the present paper, we report results of numerical investigations of the
motion of a domain wall driven by an electric current in the complete range of
parameters. The conduction electrons are treated quantum mechanically within a
lattice model. Quantum interference, spin mixing, and spin wave generation are
taken into account. The time evolution of the local magnetization is described
using the Landau-Lifshitz-Gilbert equations (LLG) in the presence of the
magnetization of the current carrying state. The electric (spin-dependent)
current in the presence of the domain wall is determined as a function of time
by using the quantum mechanical transfer matrix approach. When the Fermi
energy $E_{\rm F}$ of the conduction electrons (measured from the band edge)
is larger than the exchange energy $J_{\rm sd}$ between their spins and the
local spins of the domain wall, the state of conduction electrons is spin
mixed. Then interference effects lead to spin precession in the system of the
conduction electrons. This excites spin waves in the system of the local spins
which propagate in the direction of the current. The spin waves distort the
shape of the domain wall from the saturation configuration of the adiabatic
approximation. As a consequence, the domain wall can move beyond the
saturation displacement. However, the resulting average velocity is much
smaller than the one obtained from the spin torque only. In addition to this,
we obtain strong fluctuations of the velocity which, to the best of our
knowledge, have not been addressed before. For $E_{\rm F} < J_{\rm sd}$, the
minority spin state becomes a damped mode outside the domain wall region.  The
spin precession of the conduction electrons does not occur, and the spin mixed
state merely mimics an external magnetic field. This induces a small residual
domain wall velocity below the critical depinning current predicted by taking
into account only the spin torque mechanism. Our results strongly indicate
that the influence of spin waves cannot be neglected when interpreting
experimental data. In particular, we suggest that the small velocity of the
domain wall observed in experiments for currents well below the critical
current results from the non-adiabatic contribution of the conduction electron
spin distribution. This could be viewed as the microscopic mechanism behind
the non-adiabatic effective field that has been suggested previously
\cite{thiaville,Li04b}.

We consider a one dimensional ferromagnetic spin chain in the $x$-direction
(spins ${\bf S}_{i}$ at the sites $i$) and a fully polarized electron
propagating in the $+x$-direction coupled via s-d exchange. For the
magnetization ${\bf M}_i\propto {\bf S}_{i}$ 
\begin{eqnarray}
\frac{d{\bf M}_i}{dt}=-{\bf M}_i\times{\bf H}_{{\rm eff}i}
+ \frac{\alpha}{M_s} {\bf M}_i\times \frac{d{\bf M}_i}{dt},
\end{eqnarray}
where $\alpha$ the Gilbert damping parameter and $M_{\rm s}$ the saturation
magnetization. The effective magnetic field
\begin{eqnarray}
{\bf H}_{{\rm eff}i}&=&H_{\rm ex}\left({\bf M}_{i-1}+{\bf M}_{i+1}\right)
+ H_{K}\frac{M_{zi}}{M_s}{\bf e}_z \nonumber\\
&& \qquad- M_{yi}{\bf e}_y
- H_{\rm sd}\expect{{\bf s}}_i,
\end{eqnarray}
acts on the local spin at site $i$, with $H_{\rm ex}$ and $H_{\rm sd}$ the
exchange constant between the nearest neighbor spins and the s-d exchange
constant, respectively, and $H_{K}$ and $M_{yi}$ the anisotropy constant and
the demagnetization field, respectively. The expectation value $\expect{{\bf
    s}}_i$ is the s-d exchange field due to the polarized conduction
electrons.

In the adiabatic approximation, the conduction electrons are assumed to
propagate without reflection and their spin directions follow the local
spins. The corresponding effective magnetic field is $\propto {\bf
  M}\times ({\bf \nabla} \cdot {\bf j}_e){\bf M}$ where ${\bf j}_e$ is the
unit vector in the direction of the current \cite{Li04a}.  This is suitable
for describing a ferromagnetic metal where the Fermi wavelength is much
smaller than the system size and the width of the domain wall.

However, for the mesoscopic quantum system considered here, one must take into
account interference and spin mixing of the states of the conduction
electrons. For magnetic semiconductors, for example, the Fermi energy is of
the order of a few meV and the phase coherent length can exceed several
$\mu$m. In order to treat conduction electrons in such mesoscopic systems we
have to use numerical techniques. Without interaction, the most simple
lattice Hamiltonian (lattice parameter $a=1$) is
\begin{eqnarray}
H=-\sum_{<i,j>}c^+_ic_j-J_{\rm sd}\sum_ic_i^+\boldsymbol{\sigma}c_i
\cdot {\bf S}_i,
\end{eqnarray}
with $J_{\rm sd}>0$ and $c^+_i (c_i)$ denoting the creation (annihilation)
operator of electron at the $i$-th site. The hopping energy in the first term
is assumed to be the energy unit, and hopping is restricted to nearest
neighbors. The $\boldsymbol{\sigma}$ are the Pauli spin matrices.  By using
the transfer matrix method \cite{Pendry,Ohe03} one can calculate not only the
transmission and the reflection amplitudes of the domain wall but also the
spin resolved current carrying states $\psi_{\rm ccs}$. The expectation value
of the s-d exchange field is
\begin{eqnarray}
\expect{{\bf s}}=\bra{\psi_{\rm ccs}}
\frac{\hbar}{2}\boldsymbol{\sigma}\ket{\psi_{\rm ccs}}.
\end{eqnarray}

\begin{figure}[htb]
\begin{center}
\includegraphics[scale=1.1]{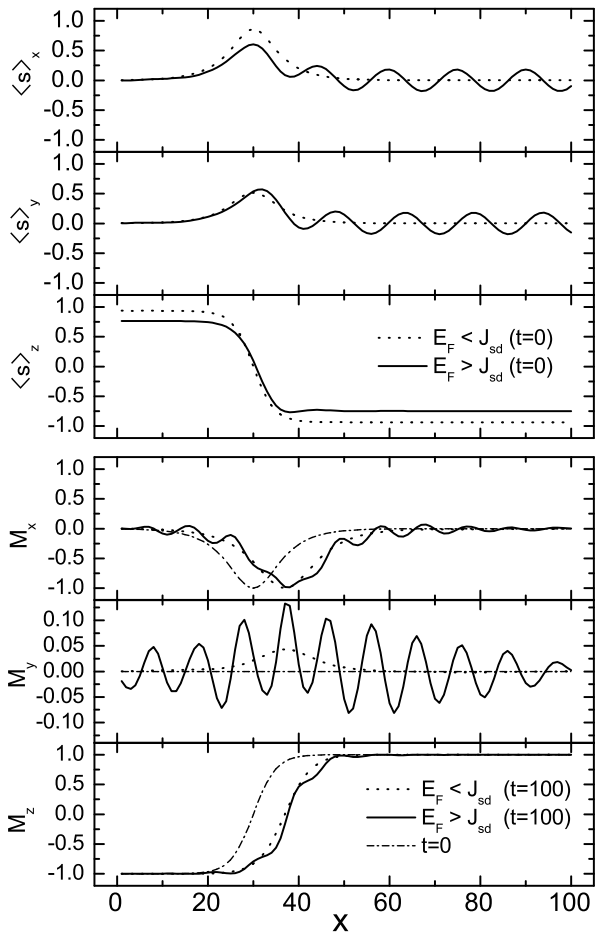}
\caption{\label{ohefig:1}
  Top: The initial s-d exchange field of the conduction electrons,
  $\expect{{\bf s}}$, calculated by the transfer matrix method. Bottom:
  The local spin configuration for $t=0$ (dashed dotted lines) and after time
  evolution ($t=100$). Parameters: $H_{\rm ex}=0.3$, s-d exchange $H_{\rm
    sd}=0.1$, anisotropy field $H_{\rm K}=0.1$; adiabatic region
  $E_{\rm F}=0.11$, $J_{\rm sd}=0.2$ (dotted lines); mesoscopic region $E_{\rm
    F}=0.21$, $J_{\rm sd}=0.2$, $\alpha=0.03$ (full lines)}
\end{center}
\end{figure}

We determine the time evolution of the magnetization according to the LLG by
taking into account this field. Both, the spin reaction torque and the
momentum transfer are included. However, for the parameters used here, it
turns out, that the momentum transfer is very small because the reflection
probability of the system is less than $10^{-2}$. Since the current carrying
state is influenced by the variation of the local spins, we have calculated
the current carrying state in each time step. The calculation proceeds as
follows. (1) An {\em ad hoc} initial ($t=0$) spin configuration of the local
spins ${\bf S}_i$ of the domain wall is assumed. (2) The s-d exchange field
$\expect{{\bf s}}_i$ is calculated for a given magnetization by the transfer
matrix method. (3) The time evolution of the local spins is calculated taking
into account $\expect{{\bf s}}_i$. (4) After a period $t=\delta t$ one obtains
a new configuration of local spins ${\bf S}_i$ from LLG. (5) Return to step
(2) and calculate $\expect{{\bf s}}_i$ by using the new configuration. Here,
we do not discuss the origin of the initial domain wall. In principle it can
be determined by a classical Monte-Carlo simulation for the given geometry of
the system. The current, and thus the time dependent electric conductance
contains important experimentally accessible information on the dynamics of
the domain wall which will be reported elsewhere \cite{Ohetbp}.  The initial
spin configuration is assumed to be a smooth function of the position, ${\bf
  S}=(S[\cosh(X_{\rm DW}-x)/W]^{-1},0,S\tanh(X_{\rm DW}-x)/W)$, where $W$ and
$X_{\rm DW}$ are the width and the center of the domain wall, respectively. We
have assumed $W=5$, $X_{\rm DW}=30$ and $S=1$. The resulting s-d exchange
field is shown in Fig.~\ref{ohefig:1}.

We consider two energy regions: the adiabatic, $E_{\rm F}<J_{\rm sd}$, and the
mesoscopic region, $E_{\rm F}>J_{\rm sd}$. In the former, the wave number of
the minority electron spins is purely imaginary outside the domain wall and
the spin is {\em almost} parallel to the local domain wall spin. Therefore,
the out-of-plane component ($y$-component) of the s-d exchange field exists
only in the vicinity of the domain wall where the conduction spin rotates in
the $(x$-$z)$-plane. In the mesoscopic case, the spin-mixed state is allowed
in entire system. Quantum interference of the spin states causes spin
precession of the conduction electrons (Fig.~\ref{ohefig:1}). This leads to
generation of spin waves in the system of the local spins.

The time evolution of the domain wall is obtained by using a 4-th order
Runge-Kutta method. The structure of the magnetization is shown for $t=100$ in
Fig.~\ref{ohefig:1}. The parameters are $H_{\rm ex}=3.0$, $H_{\rm sd}=0.1$,
$H_{\rm K}=0.1$, $M_s=1.0$ and $\alpha=0.03$. Fixed boundary conditions are
assumed at the edges of the system of the local spins. One notes the
displacement of the domain wall and that the out-of-plane component of the
domain wall is developed in both regions. In the adiabatic case, saturation of
the out-of-plane component is obtained, in agreement with previous works
\cite{Li04a}. In the mesoscopic case, one observes spin waves that have been
excited via the s-d exchange field. These cause a distortion of the domain
wall because they propagate along the current direction.

\begin{figure}[htb]
\begin{center}
\includegraphics[scale=0.87]{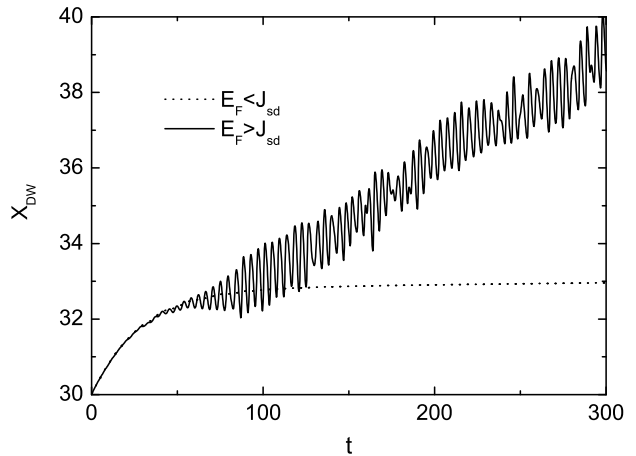}
\caption{\label{ohefig:2}
  The position of the center of the domain wall ($X_{\rm DW}$) as a function
  of time for the adiabatic (dotted line) and the mesoscopic case (full line).
  Parameters: sd-exchange $H_{\rm sd}=0.04$, anisotropy field $H_{\rm K}=0.1$,
  $\alpha=0.03$.}
\end{center}
\end{figure}

Figure~\ref{ohefig:2} shows the center of the domain wall as a function of the
time. We define the center of the domain wall as the position where $M_z=0$.
In the adiabatic region, the displacement becomes saturated after a relatively
short period of time. The saturation of the domain wall motion is due to
dissipation via damping. In the mesoscopic case, the displacement shows
fluctuating behavior that corresponds to regularly accelerating and slowing
down of the domain wall during the time evolution. This is due to the
excitation of the spin waves. They induce distortions of the wall from the
saturation configuration that is expected in the adiabatic limit. As a
consequence, after the onset of saturation due to the damping, the wall again
starts accelerating due to the distortion process before it gets slowed down
again as a result of the damping, and so on and so forth. The high-frequency
oscillations in the motion reflect standing wave patterns of spin waves
induced by the boundary conditions.

\begin{figure}[htb]
\begin{center}
\includegraphics[scale=0.84]{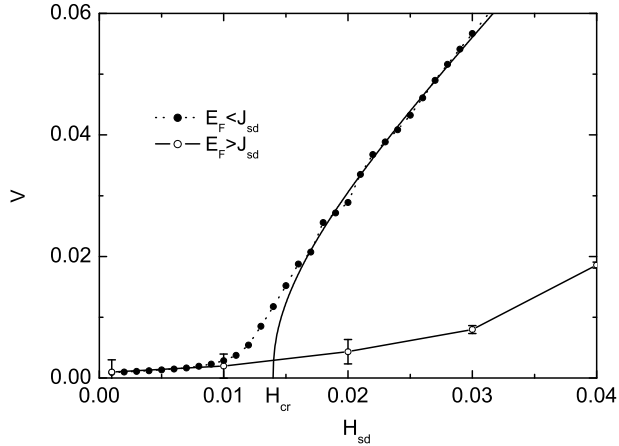}
\caption{\label{ohefig:3}
  The domain wall velocity ($V$) as a function of the s-d exchange constant
  ($H_{\rm sd}$) that is related to the amplitude of the current. The initial
  velocity is plotted for the adiabatic case ($E_{\rm F}<J_{\rm sd}$, $E_{\rm
    F}=0.11$, $J_{\rm sd}=0.2$, bullets and dotted line) while the average
  velocity is shown for the mesoscopic case ($E_{\rm F}>J_{\rm sd}$, $E_{\rm
    F}=0.21$, $J_{\rm sd}=0.2$, circles). Error bars: square root of the
  variance of the velocity. Solid line: the analytical result $\propto
  \sqrt{H_{sd}^2-H_{\rm cr}^2}$, where $H_{\rm cr}$ is the critical s-d
  exchange constant that is estimated as $H_{\rm cr}\sim 0.014$.  }
\end{center}
\end{figure}

Now we focus on the velocity of the domain wall. Figure~\ref{ohefig:3} shows
the time average as a function of the s-d coupling constant $H_{\rm sd}$. The
latter is proportional to the current. In the adiabatic case, we have plotted
the initial velocity because the motion saturates during time. It has been
predicted that the velocity is proportional to $\sqrt{H_{sd}^2-H_{\rm cr}^2}$,
where $H_{\rm cr}$ is the critical s-d exchange constant which is proportional
to the depinning current \cite{Tatara}. Our results are consistent with this
for $H_{\rm sd}>H_{\rm cr}$. However, for $H_{\rm sd}<H_{\rm cr}$ we obtain a
small but finite residual velocity as shown in Fig.~\ref{ohefig:3}. In this
region, the motion of the domain wall saturates also, but after a much longer
time than above the critical current. The residual velocity is due to the
non-adiabaticity induced by the non-adiabatic spin mixing in the state of the
conduction electrons \cite{Li04b}. The s-d exchange field and the local spin
are separated by a small angle inside the domain wall region. Thus, the domain
wall feels an effective magnetic field that corresponds to the out-of-plane
component of the s-d exchange field. The effective magnetic field induces a
displacement of the wall even below $H_{\rm cr}$ where the spin torque is
canceled out by the anisotropy field.

For the mesoscopic case, we have plotted the average velocity and the square
root of its variance because the propagation of the wall includes
fluctuations, as shown in Fig~.\ref{ohefig:2}. For $H_{\rm sd}>H_{\rm cr}$,
the velocity is small compared with the initial velocity obtained in the
adiabatic case. This discrepancy can be explained by considering the interplay
between the spin torque and the spin waves as mentioned before. In the
adiabatic limit the spin torque is the only mechanism that contributes.

In conclusion, we have numerically studied the dynamics of the domain wall
driven by a mesoscopic current. We have treated the conduction electrons
quantum mechanically. Spin-mixed states are taken into account. In the
adiabatic case, $E_{\rm F}<J_{\rm sd}$, the contribution of the spin mixed
state is equivalent to an additional external magnetic field which induces a
residual velocity of the domain wall. In the mesoscopic region, $E_{\rm
  F}>J_{\rm sd}$, the spin precession of the conduction electrons induces spin
wave excitations in the system of the local spins. We have pointed out that
the presence of these spin waves enhances the displacement. The resulting
average velocity of the domain wall is small compared with the one obtained
from the adiabatic spin torque mechanism only. In the limit of small driving
current the velocities of the adiabatic and the mesoscopic cases coincide
within the error bars. Our results suggest that non-adiabatic effects, that can
always be expected to be present in experiment, give rise to distortions of
the local spin configuration and lead to a finite velocity of the wall even
for currents below the critical value. We suggest that this is the microscopic
mechanism that accounts for the non-adiabatic contribution towards the shape
of the domain wall discussed earlier \cite{thiaville,Li04b}. This could
explain very well the results of the recent experiments where rather small
domain wall velocities have been found for currents that are an order of
magnitude below the critical current \cite{Yamaguchi,Tatara}.

An alternative explanation of the experimentally observed velocities could be
the presence of spin waves which could be due to temperature effects even for
an adiabatic system. This would lead to much smaller velocities than predicted
adiabatically (cf. Fig.~\ref{ohefig:3}). Furthermore, from
our results, the spin-wave induced velocity is predicted to show large
fluctuations that are consistent with the experimental observations
\cite{Yamaguchi}.

Finally, we want to point out that our calculations have been done for a
general abstract model. By suitably adjusting the parameters, and by
generalizing the model Hamiltonian, for instance by including more than one
conduction channel, our method may be able to contribute to better
understanding a variety of different experimental situations ranging from
ferromagnetic nanowires to magnetic semiconductor nanostructures. In addition,
since we treat the conduction electrons fully quantum mechanically, mesoscopic
interference effects in the current as a function of time can in principle be
expected --- especially for magnetic semiconductor nanowires
\cite{ohno,ruester} --- that could be signaled in the dynamics of the
domain wall. Whether or not this will be the case is an open question and the
subject of future research.

\acknowledgements{The authors are grateful to H.-P. Oepen, M. Yamamoto and T.
  Ohtsuki for valuable discussions. The work has been supported by the
  Deutsche Forschungsgemeinschaft via the SFB 508 of the Universit\"at
  Hamburg, and by the European Union via the Marie-Curie-Network
  MCRTN-CT2003-504574.}


\end{document}